\pdfoutput=1
\documentclass[aps,prb,superscriptaddress,reprint]{revtex4-2}
\usepackage{graphicx}
\usepackage{color}
\usepackage{amsmath,amsthm,amssymb,amsfonts}
\usepackage{physics, ulem, cancel}
\usepackage{physics2}
\usephysicsmodule{ab, ab.braket}
\usepackage{cases}
\usepackage{mathrsfs}
\usepackage{mathtools,bm}
\usepackage{subfigure,here}
\usepackage{comment}
\usepackage{hyperref}
\usepackage{simpler-wick}
\definecolor{darkslateblue}{RGB}{75,60,220}
\definecolor{darkred}{RGB}{200,0,0}
\hypersetup{
citecolor = darkslateblue,
colorlinks = true,
urlcolor = darkslateblue,
linkcolor = darkred
}

\usepackage{ulem}

\begin{document}

\title{
    Designing XY and Dzyaloshinskii--Moriya couplings in Majorana Cooper pair boxes}

\author{Manato Teranishi}
\affiliation{Department of Physics, Nagoya University, Nagoya 464-8602, Japan}
\author{Shintaro Hoshino}
\affiliation{Department of Physics, Chiba University, Chiba 263-8522, Japan}
\author{Ai Yamakage}
\affiliation{Department of Physics, Nagoya University, Nagoya 464-8602, Japan}

\date{\today}

\begin{abstract}
    We theoretically study how to design spin couplings in networks of Majorana Cooper pair boxes (MCBs) connected by multiple normal-metal leads.
    The inter-box interaction is generated by the conduction-electron-mediated Ruderman--Kittel--Kasuya--Yosida (RKKY) interaction.
    We show that the connectivity of Majoranas to the leads enables arbitrary types of couplings.
    As concrete examples, we show the realization of the XY and Dzyaloshinskii--Moriya (DM) couplings, which are difficult to implement in previously proposed MCB-based schemes.
    The sign and magnitude of the couplings can be tuned continuously via gate-controlled tunneling amplitudes.
    These results establish MCBs as a versatile platform for engineered quantum spin systems.
\end{abstract}

\maketitle

\section{Introduction}
\label{sec1}

Quantum spin systems serve as a fundamental platform for understanding diverse quantum phenomena, such as magnetism and topological phases.
Recent advances in quantum simulation techniques have enabled the realization of various quantum spin models using ultracold atomic systems, such as the Ising model
\cite{Ising1,Ising2,Ising2p,Ising3,Ising4,Ising5,Ising6,Ising7,Ising7m,Ising8,Ising8p,Ising9}, the XY model \cite{ex_XY1,ex_XY2,ex_XY3}, the XXZ model \cite{XXZ1,XXZ2,XXZ3,XXZ4}, and the XYZ model \cite{XYZ1,XYZ2}.
Moreover, theoretical studies propose the realizations of Dzyaloshinskii–Moriya (DM) interactions \cite{DM1,DM2,DM3,DM4,DM5}, spin-1 models \cite{spin1Rydberg1,spin1Rydberg2}, and the Kitaev model \cite{Kitaev1,Kitaev2}.

Parallel to these developments, Majorana Cooper pair boxes (MCBs), mesoscopic superconducting islands hosting Majoranas, have emerged as a promising alternative platform for realizing quantum spin systems \cite{SimulatingSpinMF,sagi_spin_2019,Ebisu_2019_SUSY,FractonMF,MCB_review2020,Tsvelik2020,YangLeeMF}.
An MCB comprises multiple topological superconducting nanowires with localized Majoranas at their ends \cite{Majorana_Sau_PRL,Majorana_Alicea_PRB,Majorana_Oreg_PRL}.
Due to the large charging energy, an effective spin-1/2 degree of freedom is defined on a MCB \cite{beriTopologicalKondoEffect2012,MCB_Plugge_2017}.
Therefore, constructing a network of MCBs is a central strategy for designing quantum spin systems with MCBs.
This scheme realizes a wide variety of spin models, including the XYZ model and the transverse-field Ising model \cite{sagi_spin_2019}.
Furthermore, it has been suggested that a two-dimensional arrangement of MCBs can lead to the realization of the Kitaev model, offering a platform for quantum spin liquids \cite{SimulatingSpinMF,sagi_spin_2019,Tsvelik2020}.
Intriguingly, proposals for exotic systems such as the spin-1 Blume--Capel model with supersymmetry \cite{Ebisu_2019_SUSY}, models exhibiting fracton phases \cite{FractonMF}, and non-Hermitian models hosting Yang--Lee anyons \cite{YangLeeMF}, highlight the versatility of MCBs as a platform for engineering unconventional quantum spin systems.

However, existing protocols for MCB-based quantum spin systems have primarily focused on standard symmetric couplings, such as Ising or Heisenberg couplings, and have not yet established methods for generating arbitrary spin--spin couplings.
Specifically, designs for XY-type couplings and asymmetric spin couplings, such as the DM coupling, remain unexplored.
In this work, we address this challenge by utilizing the Ruderman--Kittel--Kasuya--Yosida (RKKY) framework \cite{PhysRev.96.99, Kasuya1956, PhysRev.106.893, Tsvelik2020} with various connection patterns of the leads.
We elucidate that the freedom in the wiring pattern is the key to engineering diverse spin Hamiltonians and demonstrate the realizations of the XY coupling and the Heisenberg interaction with the DM coupling.

This paper is organized as follows.
In Sec.~\ref{2_model}, we introduce a general model consisting of two MCBs connected via normal metal leads and derive its low-energy effective Hamiltonian.
In Sec.~\ref{3_RKKY}, we derive the general form of the RKKY interaction mediating the coupling between the effective spins.
We then demonstrate the engineering of specific spin couplings, including the XY coupling and the Heisenberg coupling with the DM coupling, in Secs.~\ref{sec_XY} and \ref{sec_HDM}, respectively.
In Sec.~\ref{4_ss}, we calculate the spin susceptibility of the leads for two cases: infinite-length leads and finite-length leads.
In Sec.~\ref{estimate}, we estimate the energy scale of the RKKY interaction using realistic experimental parameters.
Finally, Sec.~\ref{5_Summary} is devoted to the summary and discussion of future perspectives.
\section{Model}
\label{2_model}

\begin{figure*}
    \centering
    \subfigure[]{
        \includegraphics[scale=0.19]{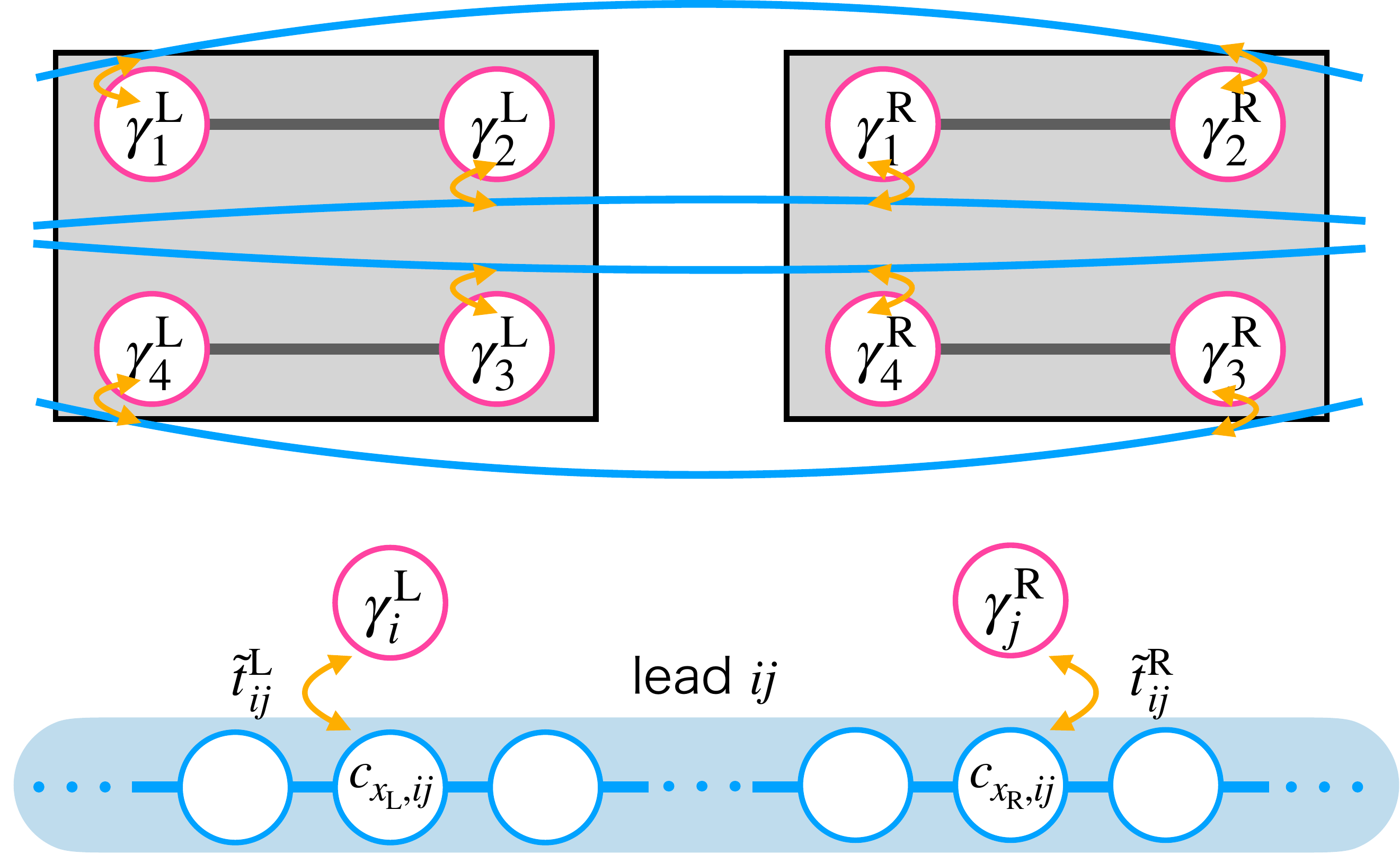}\label{fig:system_inf}}
    \hspace{1cm}
    \subfigure[]{
        \includegraphics[scale=0.19]{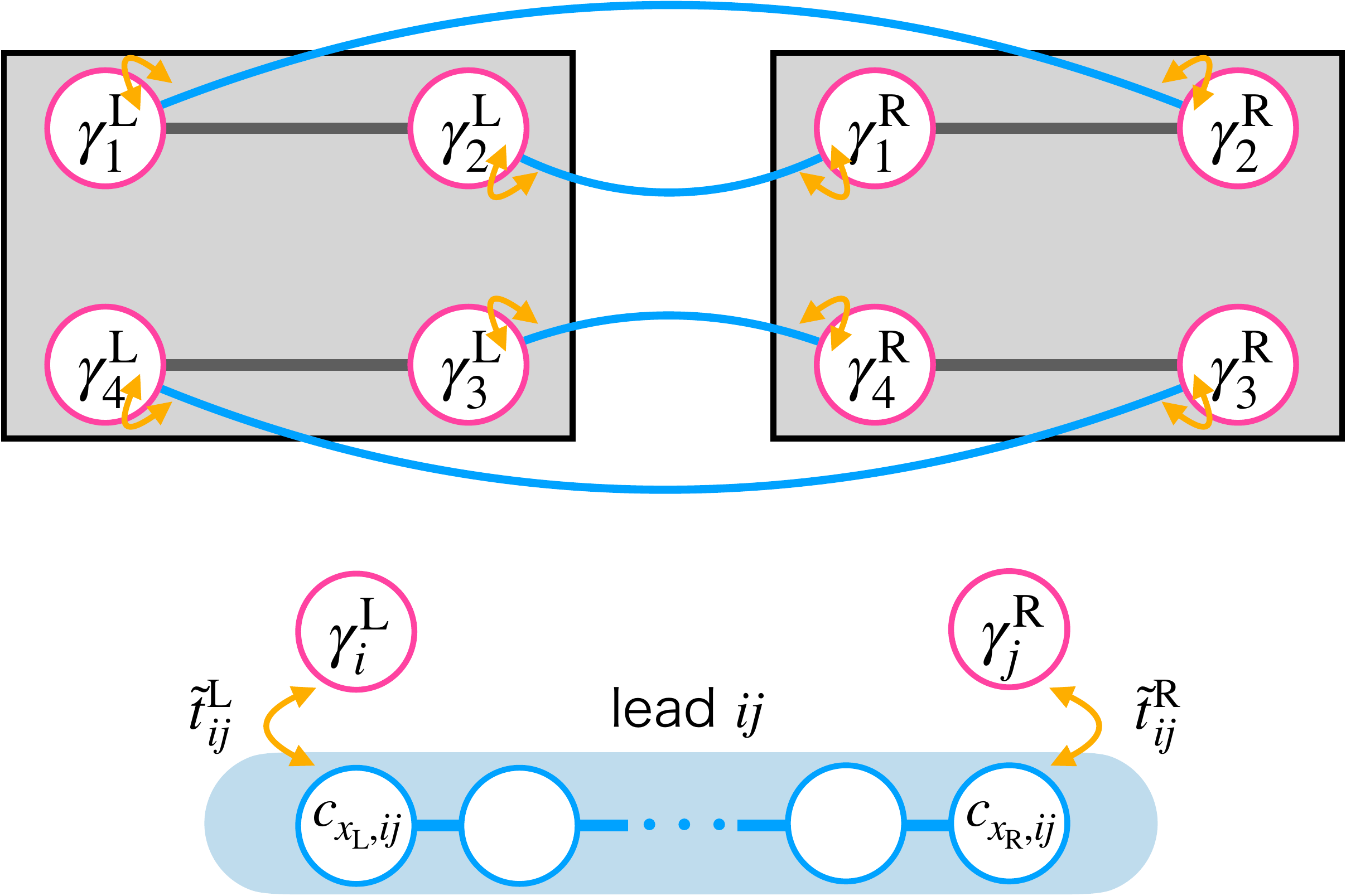}\label{fig:system_fin}}
    \caption{
        Schematic setup for the two-spin system formed by two MCBs.
        The left (L) and right (R) boxes are MCBs with charging energy $E_{\rm c}^{\alpha}$ ($\alpha = {\rm L, R}$), where each of the four Majoranas (pink circles) is connected to leads (blue wires). 
        Yellow double-headed arrows denote the tunnel coupling between the Majoranas and the leads. 
        Panels (a) and (b) show the cases of infinite and finite leads, respectively.}
    \label{fig:system_0}
\end{figure*}

We consider two MCBs connected by multiple leads, as illustrated in Fig.~\ref{fig:system_0}.
Each MCB consists of two semiconducting nanowires with Rashba spin--orbit coupling placed on an $s$-wave superconductor, under a magnetic field.
The nanowires are assumed to be in the topological superconducting phase hosting Majorana zero modes at the ends, and the corresponding Majorana operators are denoted by $\gamma_i^\alpha\ (i=1,2,3,4)$, where the superscript $\alpha=\mathrm{L}, \mathrm{R}$ indicates the MCB at the left or right side, respectively.
The mesoscopic size of the MCBs induces a large charging energy, which restricts the system to a fixed-parity sector, allowing each MCB to be treated as an effective spin-1/2 system.
In this section, we describe the Hamiltonian of the system and then derive the effective Hamiltonian.

\subsection{Two-box model}
We consider a setup in which two MCBs, labeled L and R, are connected via multiple leads.
We assume that each of the four Majoranas in one MCB is connected to all four Majoranas in the other, resulting in a total of 16 leads.
The Hamiltonian $\mathcal H$ is composed of the kinetic energy of the leads $\mathcal{H}_{\rm lead}$, tunneling between leads and Majoranas $\mathcal{H}_{\rm tun}$, charging energy of MCB $\mathcal{H}_{\rm c}$, and hybridization of Majoranas $\mathcal{H}_{\rm Z}$,
\begin{align}
    \mathcal{H} = \mathcal{H}_{\mathrm{lead}} +\mathcal{H}_{\mathrm{tun}}+ \mathcal{H}_{\mathrm{c}} +\mathcal{H}_{\mathrm{Z}}. \label{hamiltonian}
\end{align}
The first term is the lead Hamiltonian as a one-dimensional tight-binding model as follows:
\begin{align}
    \mathcal{H}_{\mathrm{lead}}
     & = \sum_{i,j=1}^4
    \sum_{r}
    \Bigg\lbrack \varepsilon_0 c_{r,ij}^{\dag} c_{r,ij}
    -t \qty(c_{r,ij}^{\dag} c_{r+1,ij} +\mathrm{h.c.} )\Bigg\rbrack, \label{H_lead}
\end{align}
where $\varepsilon_0$ is the on-site energy, $t$ is the nearest-neighbor hopping energy, and $c_{r,ij}^{\dag}$ and $c_{r,ij}$ denote the spinless electron creation and annihilation operators at a site $r$ in the lead labeled by $ij$, respectively.
Note that only the relevant spin projection of electrons, which is regarded as a spinless fermion, coupled to the Majoranas, is taken into account \cite{spinlessPRL2007, spinlessPRB2020}.
We consider two types of lead configurations: infinite-length leads with periodic boundary conditions (PBC), as shown in Fig.~\ref{fig:system_inf}, and finite-length leads with open boundary conditions (OBC), as shown in Fig.~\ref{fig:system_fin}.

The second term $\mathcal H_{\rm tun}$, which is the hybridization between the Majoranas and the electrons in the leads, is given by the following tunneling Hamiltonian \cite{Fu_Majorana_2010, beriTopologicalKondoEffect2012}
\begin{align}
    \mathcal{H}_{\mathrm{tun}}
     & =\sum_{i,j=1}^4
    \qty(
    \tilde{t}_{ij}^{\rm L} \gamma_{i}^{\rm L} c_{x_{\rm L},ij}
    +
    \tilde{t}_{ij}^{\rm R} \gamma_{j}^{\rm R} c_{x_{\rm R},ij}
    )
    +\mathrm{h.c.},
    \label{Htun}
\end{align}
where $\tilde{t}_{ij}^{\alpha}=t_{ij}^{\alpha}e^{i\phi_{\alpha,ij}}$ ($t_{ij}^\alpha > 0$) denotes the tunneling amplitude between $\gamma_i^\alpha$ and $c_{x_\alpha,ij}\ (c_{x_\alpha,ij}^\dagger)$ in the lead $ij$ that connects $\gamma_i^{\rm L}$ and $\gamma_j^{\rm R}$, and $\phi_{\alpha,ij}$ is the relative phase difference between the phase of the superconducting order parameter of each MCB and each lead.
As shown in the bottom panel of Fig.~\ref{fig:system_0}, the electron is hybridized with the Majoranas at $x_{\rm L}$ on the left MCB and $x_{\rm R}$ on the right MCB, respectively.
Here, the $i$th Majorana operator in the MCB labeled $\alpha$ satisfies
\begin{align}
    \gamma^{\alpha}_i=\gamma_i^{\alpha\dagger},\ \acomm{\gamma^{\alpha}_{i}}{\gamma^{\beta}_{j}}=2\delta_{ij} \delta_{\alpha\beta}.
\end{align}

The third term $\mathcal H_{\rm c}$ denotes the charging energies of the MCBs and is given by
\begin{align}
    \mathcal{H}_{\mathrm{c}}
     =\sum_{\alpha={\rm L},{\rm R}}E_{\mathrm{c}}^\alpha \qty(\mathcal{N}^\alpha - N_{\mathrm{g}}^\alpha)^2, \label{H_charging}
\end{align}
where $E_{\mathrm{c}}^\alpha$ is the charging energy of each MCB, $\mathcal{N}^\alpha$ is the total electron number operator in each MCB, and $N_{\mathrm{g}}^\alpha$ is the gate charge, which is tuned externally by applying a backgate voltage.

The last term $\mathcal{H}_{\mathrm{Z}}$ describes the overlap of Majorana wavefunctions, and is written as follows \cite{Kitaev_2001}:
\begin{align}
    \mathcal{H}_{\mathrm{Z}}
     = i\sum_{\alpha={\rm L},{\rm R}} \sum_{i, j=1}^4 \tau_{ij}^\alpha \gamma_{i}^\alpha \gamma_{j}^\alpha, \label{Zeeman}
\end{align}
where the coupling constant is antisymmetric as $\tau_{ij}^\alpha = -\tau_{ji}^\alpha$.
The subscript ``Z'' derives from the fact that it gives rise to a Zeeman term in the effective spin system.
The magnitude and sign of this term can be controlled by tuning the nanowire length and chemical potential.

\subsection{Mapping from Majoranas to spin degrees of freedom}
In this paper, we focus on the case where the charging energy is the largest energy scale in the system, and the fermion parity for each MCB is fixed as \cite{sagi_spin_2019}
\begin{align}
    \mathcal{P}^{\alpha}=(i\gamma^{\alpha}_1\gamma^{\alpha}_2)(i\gamma^{\alpha}_3\gamma^{\alpha}_4)=\pm 1. \label{FPconstraint}
\end{align}
As a result, the degrees of freedom of the fixed-parity subspace are reduced from $2^2=4$ to 2, which is represented as a spin-1/2
\begin{align}
    S_i^{\alpha} \coloneqq \frac{1}{2}\sum_{j,k=1}^3 \varepsilon_{ijk} L_{jk}^\alpha = \mathcal P^{\alpha} L_{i4}^{\alpha}, \label{SpinOP}
\end{align}
where $\varepsilon_{ijk}$ is the Levi-Civita symbol, and $L^{\alpha}_{jk} \coloneqq -i\gamma^{\alpha}_j\gamma^{\alpha}_k/2$.
The spin operators $S_1$, $S_2$, and $S_3$ correspond to $S_x$, $S_y$, and $S_z$, respectively.

\subsection{Effective Hamiltonian}
We consider the low-energy effective theory given by the Schrieffer-Wolff transformation
\cite{PhysRev.149.491} within the second-order perturbation of Eq.~\eqref{Htun} when the tunneling amplitudes are sufficiently small compared to the charging energy ($t_{ij}^{\alpha}/E_{\mathrm{c}}^\alpha \ll 1$).
The effective Hamiltonian is given by
\begin{align}
    \mathcal{H}_{\rm eff} & = \mathcal{H}_{\rm lead} + \mathcal{H}' + \mathcal{H}_{\rm scatt} +\mathcal{H}_{\mathrm{Z}},
    \label{Heff}
\end{align}
with
\begin{align}
    \mathcal H'
     & =
    2\sum_{i \neq j} \sum_{k,l=1}^4
    \biggl[
        \frac{(\tilde t_{ik}^{\rm L})^* \tilde t_{jl}^{\rm L}}{E_{\mathrm{c}}^{\rm L}}
        \gamma_j^{\rm L} \gamma_i^{\rm L}
        c_{x_{\rm L}, ik}^\dag
        c_{x_{\rm L}, jl}
    \notag \\&\quad
        +
        \frac{(\tilde t_{ki}^{\rm R})^*
            \tilde t_{lj}^{\rm R}}{E_{\mathrm{c}}^{\rm R}}
        \gamma_j^{\rm R} \gamma_i^{\rm R}
        c_{x_{\rm R}, ki}^\dag
        c_{x_{\rm R}, lj}
        \biggr],
\end{align}
where the sum $\sum_{i \neq j}$ runs from 1 to 4.
The $i = j$ term corresponds to a potential scattering $\mathcal H_{\mathrm{scatt}}$.
In the lowest-order perturbation calculation of the RKKY interaction in Sec.~\ref{3_RKKY}, this term does not contribute spin degrees of freedom; therefore, it is neglected in this paper.

Next, we rewrite the effective Hamiltonian Eq.~\eqref{Heff} in spin representation by Eq.~\eqref{SpinOP} as
\begin{align}
    \mathcal{H}' =
    \sum_{i<j}\sum_{k,l}
    \Big(
    T^{\rm L}_{ik,jl} L^{\rm L}_{ij} J^{\rm L}_{ik,jl}
    + T^{\rm R}_{ki,lj} L^{\rm R}_{ij} J^{\rm R}_{ki,lj}
    \Big),
\end{align}
where $T_{ij,kl}^\alpha \coloneqq 4 t_{ij}^\alpha t_{kl}^\alpha / E_{\mathrm{c}}^\alpha$ is a coupling constant defined for a pair of leads $ij$ and $kl$.
$J^{\alpha}_{ij,kl}$ is defined from the electron operators in the leads as
\begin{align}
    J^{\alpha}_{ij,kl}
    \coloneqq -i
    e^{-i(\phi_{\alpha,ij}-\phi_{\alpha,kl})} c_{x_\alpha,ij}^{\dagger} c_{x_\alpha,kl}
    +\mathrm{h.c.},
\end{align}
which is interpreted as the generators of the SO(16) symmetry group corresponding to the conduction-electron spin operator.

The bilinear part of the Majorana operators is regarded as the Zeeman term
\begin{align}
    \mathcal{H}_{\mathrm{Z}}
    =\sum_{\alpha}\sum_{i}
    B_{i}^\alpha S_{i}^\alpha , \label{HZeeman}
\end{align}
where $B_i^\alpha=-4(\tau_{jk}^\alpha+\mathcal{P}^\alpha\tau_{i 4}^\alpha)$.
\section{RKKY interaction}
\label{3_RKKY}

First, we derive the RKKY interaction $\mathcal{H}_{\mathrm{RKKY}} = \sum_{i<j} \sum_{i'<j'} \mathcal J_{ij,i'j'} L_{ij}^{\rm L} L_{i'j'}^{\rm R}$ for our model.
The RKKY interaction can be obtained from the second-order correction to the free energy as follows \cite{RKKYinGraphene, RKKYfreeEnergy}:
\begin{align}
    \mathcal J_{ij,i'j'}
     &
    = -T\sum_{klk'l'}
    T_{ik,jl}^{\rm L} T_{k' i', l' j'}^{\rm R}
    \int_0^{1/T} d\tau_1
    \int_0^{1/T} d\tau_2
    \notag \\&\quad\times
    \ev{T_\tau J_{ik, jl}^{\rm L}(\tau_1) J_{k'i',l'j'}^{\rm R}(\tau_2)}, \label{Jijij}
\end{align}
where the average $\ev{\cdots} = \mathrm{tr} (e^{-\mathcal H_{\mathrm{lead}}/T} \cdots)/\mathrm{tr} (e^{-\mathcal{H}_{\mathrm{lead}}/T})$ is taken for the connected terms.
By performing the integration over imaginary time, we obtain the following result (see Appendix~\ref{App_RKKY} for details),
\begin{align}
    \mathcal{J}_{ij,i'j'} & =
    \mathcal{T}_{ij'}\mathcal{T}_{ji'}
    \widetilde{\chi}_{ij',ji'}
    -
    \mathcal{T}_{ii'}\mathcal{T}_{jj'}
    \widetilde{\chi}_{ii',jj'},
    \label{Jij}
\end{align}
where we define the product of the tunneling amplitudes between Majoranas in lead $ij$ as $\mathcal{T}_{ij}\coloneqq 4t^{\rm L}_{ij}t^{\rm R}_{ij}/\sqrt{E_{\rm c}^{\rm L} E_{\rm c}^{\rm R}}$.
Moreover,
\begin{align}
    \widetilde{\chi}_{ij,kl} & \coloneqq
    2\mathrm{Re}\qty[e^{i(\Delta\phi_{ij}-\Delta\phi_{kl})} \chi_{ij,kl}], \label{chi_til}
\end{align}
where
\begin{align}
    \chi_{ij,kl} & = -T\int_0^{1/T} \dd \tau_1 \int_0^{1/T} \dd \tau_2 \nonumber              \\
                 & \quad\times \expval{T_\tau c_{x_n,ij}(\tau_1) c_{x_m,ij}^{\dagger}(\tau_2)
        c_{x_m,kl}(\tau_2) c_{x_n,kl}^{\dagger}(\tau_1)},
    \label{chi_general}
\end{align}
corresponds to the static spin susceptibility defined by the two leads $ij$ and $kl$, and $\Delta\phi_{ij}\coloneqq \phi_{{\rm L},ij}-\phi_{{\rm R},ij}$.

Finally, by rewriting $\mathcal{H}_{\rm RKKY}$ from the operator $L_{ij}^\alpha$ to the spin operator $S_{i}^\alpha$ using Eq.~\eqref{SpinOP}, we obtain the RKKY Hamiltonian between the spins at the left and right MCBs,
\begin{align}
    \mathcal{H}_{\mathrm{RKKY}} = \sum_{i,i'} \mathcal{J}_{ii'} S_i^{\rm L} S_{i'}^{\rm R}.\label{HRKKY2}
\end{align}
Exploiting the antisymmetry of the original coupling constant, $\mathcal{J}_{jk,j'k'} = -\mathcal{J}_{kj,j'k'} = -\mathcal{J}_{jk,k'j'}$, the resultant coupling constant $\mathcal{J}_{ii'}$ is explicitly expressed as follows:
\begin{align}
    \mathcal{J}_{ii'} & =
    \frac{1}{4}\varepsilon_{ijk}\varepsilon_{i'j'k'}\mathcal{J}_{jk,j'k'}
    + \frac{1}{2}\varepsilon_{i'j'k'} \mathcal{P}^{\rm L} \mathcal{J}_{i4,j'k'}
    \nonumber                                                                                        \\
                      & \quad+ \frac{1}{2}\varepsilon_{ijk} \mathcal{P}^{\rm R} \mathcal{J}_{jk,i'4}
    + \mathcal{P}^{\rm L}\mathcal{P}^{\rm R} \mathcal{J}_{i 4,i' 4},
    \label{eq:J}
\end{align}
where summation over repeated indices ($j, k, j', k' \in \{1, 2, 3\}$) is implied.

The coupling constant Eq.~\eqref{eq:J} can be controlled by tuning the tunneling amplitude via gate voltages, the susceptibility of the leads by adjusting their lengths or chemical potentials, and the phase difference between the superconducting order parameters by driving a Josephson current.

\begin{figure}
    \centering
    \subfigure[]{
        \includegraphics[scale=0.2]{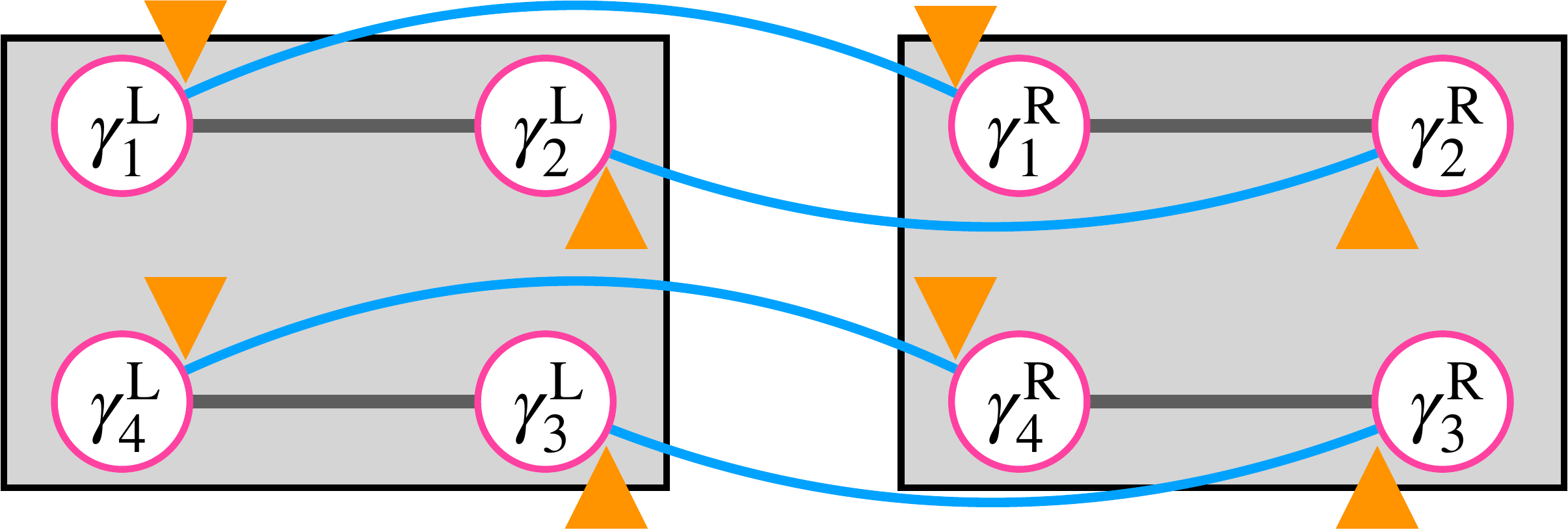}\label{fig:setupXY}}
    \subfigure[]{
        \includegraphics[scale=0.2]{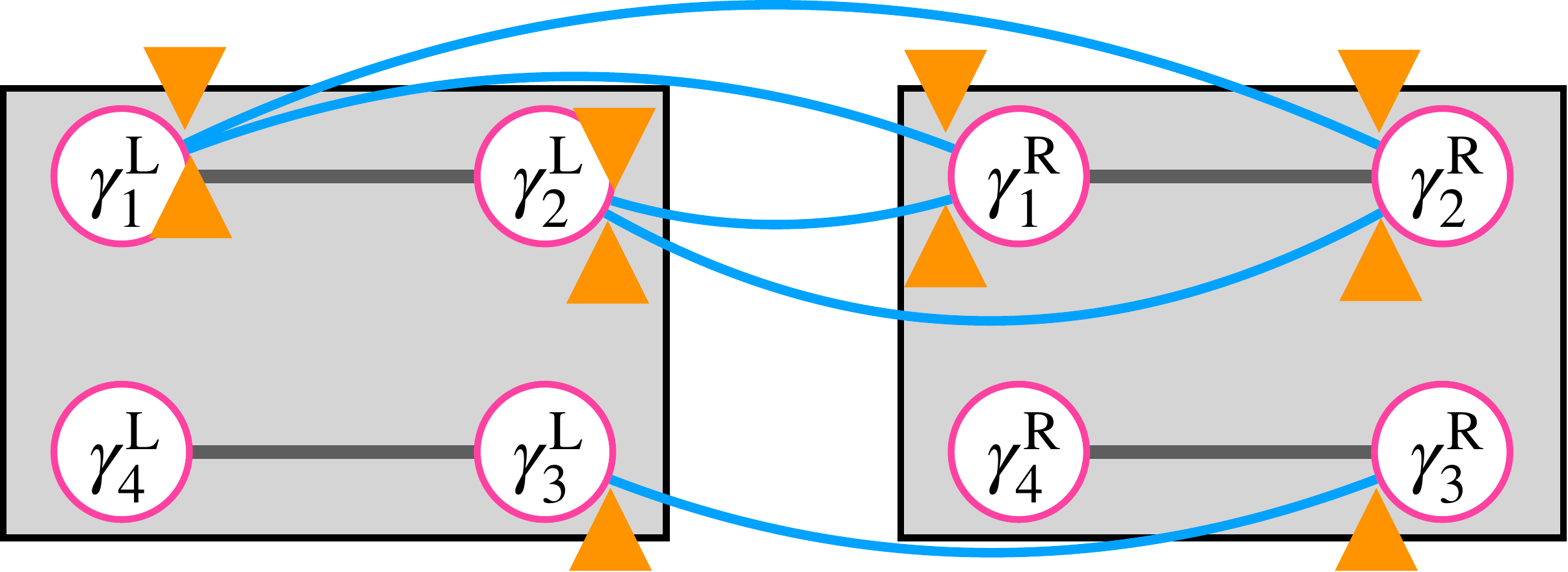}\label{fig:setupDM}}
    \caption{
        Schematic setups for realizing (a) the XY coupling and (b) the Heisenberg + DM coupling.
        The orange triangles indicate the externally applied gate voltages used to modulate the tunneling amplitudes.
    }
    \label{fig:setupXYDM}
\end{figure}

In the following, we apply the theory to the realization of the XY and the Heisenberg + DM couplings.
    We provide a specific wiring configuration shown in Fig.~\ref{fig:setupXYDM}, which serves as the minimal configuration to achieve these target interactions.
    In this setup, the gate voltages at the positions marked with orange triangles in Fig.~\ref{fig:setupXYDM}, which control the tunneling amplitudes, are treated as the only externally modulated parameters.

\subsection{XY coupling}\label{sec_XY}
We consider the XY-type spin coupling given by
\begin{align}
    \mathcal{H} = J \sum_{\mu=x,y} S^{\rm L}_{\mu} S^{\rm R}_{\mu},
\end{align}
utilizing a configuration with four leads, $ij=$ 11, 22, 33, and 44, as illustrated in Fig.~\ref{fig:setupXY}.
In this case, $\mathcal T_{ij}=0$ for $i\neq j$, and the diagonal components of the coupling constants in Eq.~\eqref{eq:J} satisfy $\mathcal J_{ii}= -\sum_{j<k}\abs{\varepsilon_{ijk}}\mathcal{T}_{jj}\mathcal{T}_{kk} \widetilde{\chi}_{jj,kk}- \mathcal{P}^{\rm L} \mathcal{P}^{\rm R} \mathcal{T}_{ii}\mathcal{T}_{44} \widetilde{\chi}_{ii,44}$.
This expression also clarifies the role of lead 44. If only the leads 11, 22, and 33 are connected, finite $\mathcal{J}_{xx}$ and $\mathcal{J}_{yy}$ are generically accompanied by a finite $\mathcal{J}_{zz}$. Lead 44 provides the additional independent contribution required to cancel $\mathcal{J}_{zz}$ while retaining the finite transverse exchange components. Therefore, this setup constitutes the minimal configuration for realizing a pure XY coupling.
We now examine how the RKKY interaction $\mathcal J$ can be controlled via $\mathcal T$.
Imposing the condition $\mathcal{J}_{zz}=0$ for the XY coupling yields
\begin{align}
    \tilde{\mathcal{T}}_{33}
    =-\mathcal{P}^{\rm L} \mathcal{P}^{\rm R}
    \frac{\widetilde{\chi}_{11,22}}{\widetilde{\chi}_{33,44}}
    \tilde{\mathcal{T}}_{11}\tilde{\mathcal{T}}_{22}, \label{XY_T33}
\end{align}
where $\tilde{\mathcal{T}}_{ii}\coloneqq\mathcal{T}_{ii}/\mathcal{T}_{44}\ (i=1,2,3)$ are normalized parameters that can be tuned externally.
Since the product of the tunneling amplitudes $\mathcal{T}_{ii}$ is always a positive real number, the product of the fermion parities and the spin susceptibilities must satisfy the sign condition: $\mathrm{sgn} (\mathcal{P}^{\rm L} \mathcal{P}^{\rm R} \widetilde{\chi}_{11,22}/\widetilde{\chi}_{33,44}) = -1$.
If $+1$ is chosen, it means there exists no $\mathcal{T}_{33}$ satisfying $\mathcal{J}_{zz}=0$.
Substituting this into the expressions for $\mathcal{J}_{xx}$ and $\mathcal{J}_{yy}$, we obtain
\begin{align}
    \tilde{\mathcal{J}}_{xx}
     & =\tilde{\mathcal{T}}_{11} \tilde{\mathcal{T}}_{22}^2 - \tilde B_x \tilde{\mathcal{T}}_{11}, \label{XY_Jx}
    \\
    \tilde{\mathcal{J}}_{yy}
     & =\tilde A_y \tilde{\mathcal{T}}_{11}^2 \tilde{\mathcal{T}}_{22} - \tilde B_y \tilde{\mathcal{T}}_{22}, \label{XY_Jy}
\end{align}
where $\tilde{\mathcal{J}}_{ii}\coloneqq\mathcal{J}_{ii}/(A_x \mathcal{T}_{44}^2)$ are the dimensionless coupling constants, and we introduce the normalized parameters $\tilde A_y \coloneqq A_y/A_x$ and $\tilde B_i \coloneqq B_i/A_x$.
The parameters $A_i$ and $B_i$ are defined as
\begin{align}
    A_i \coloneqq \mathcal{P}^{\rm L} \mathcal{P}^{\rm R} \sum_{j<k}\abs{\varepsilon_{ijk}} \frac{\widetilde{\chi}_{11,22}\widetilde{\chi}_{jj,kk}}{\widetilde{\chi}_{33,44}}, \
    B_i \coloneqq \mathcal{P}^{\rm L} \mathcal{P}^{\rm R} \widetilde{\chi}_{ii,44},
\end{align}
where the index $i \in \{1, 2, 3\}$ corresponds to $\{x, y, z\}$ on the left-hand sides.
These parameters depend only on the fermion parities of each MCB and the susceptibilities of the leads.
They are, therefore, intrinsic to the system and are treated as constants in this section.

Solving the isotropic condition with $J=\mathcal{J}_{xx}=\mathcal{J}_{yy}$, we obtain $\tilde{\mathcal{T}}_{ii}$ as a function of the normalized coupling constant $\tilde J\coloneqq J/(A_x \mathcal{T}_{44}^2)$.
Although the general solution is complicated (see Appendix~\ref{App_XY}), imposing the simplified conditions $\tilde B_x = \tilde A_y \tilde B_y$ and $\tilde A_y, \tilde B_y <0$ reduces the problem to a quadratic equation, yielding simple analytical solutions:
\begin{align}
    \tilde{\mathcal T}_{11}
     & = \sqrt{\qty(\frac{\tilde J}{2 \tilde A_y \tilde B_y})^2 + \frac{\tilde B_y}{\tilde A_y}}
    - \frac{\tilde J}{2\tilde A_y \tilde B_y}, \label{XY_T11}                                    \\
    \tilde{\mathcal{T}}_{22}
     &
    =
    \sqrt{\ab(\frac{\tilde J}{2 \tilde B_y})^2 + \tilde A_y \tilde B_y} - \frac{\tilde J}{2 \tilde B_y}.
    \label{XY_T22}
\end{align}

\begin{figure}
    \centering
    \includegraphics[width=1\linewidth]{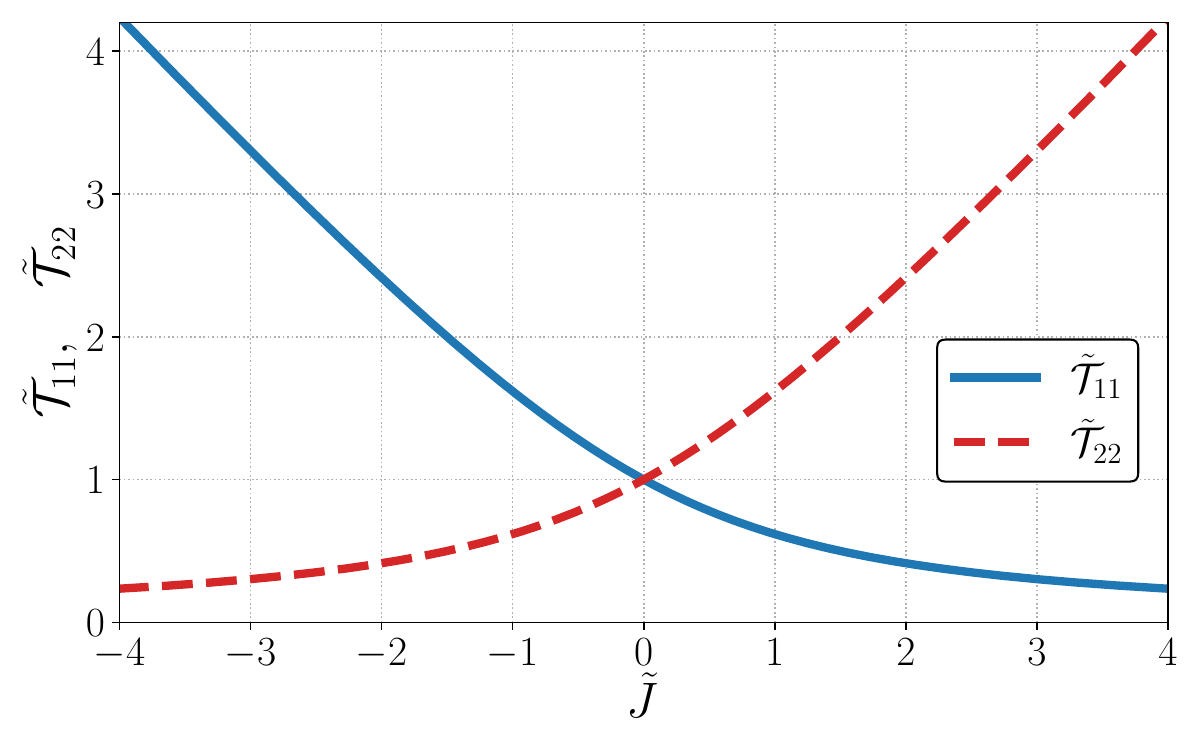}
    \caption{Tunneling amplitudes realizing the XY coupling. $\tilde{\mathcal{T}}_{11}$ (solid line) and $\tilde{\mathcal{T}}_{22}$ (dotted line) as a function of $\tilde{J}$ obtained from Eqs.~\eqref{XY_T11} and \eqref{XY_T22}, for $(\tilde B_x, \tilde A_y, \tilde B_y)=(1,-1,-1)$.
    }
    \label{fig:XY1}
\end{figure}

Figure~\ref{fig:XY1} shows the solutions $\tilde{\mathcal{T}}_{11}$ (solid blue line) and $\tilde{\mathcal{T}}_{22}$ (dashed red line) as a function of the target coupling constant $\tilde J$.
Since the parameters $A_i$ and $B_i$ are expected to be of the same order of magnitude, we set $(\tilde B_x, \tilde A_y, \tilde B_y)=(1,-1,-1)$ for example.
The plot shows that the sign of the interaction can be continuously tuned between ferromagnetic and antiferromagnetic regimes simply by adjusting the tunneling amplitudes.

\subsection{Heisenberg + DM coupling}\label{sec_HDM}
\begin{figure}
    \centering
    \includegraphics[width=1\linewidth]{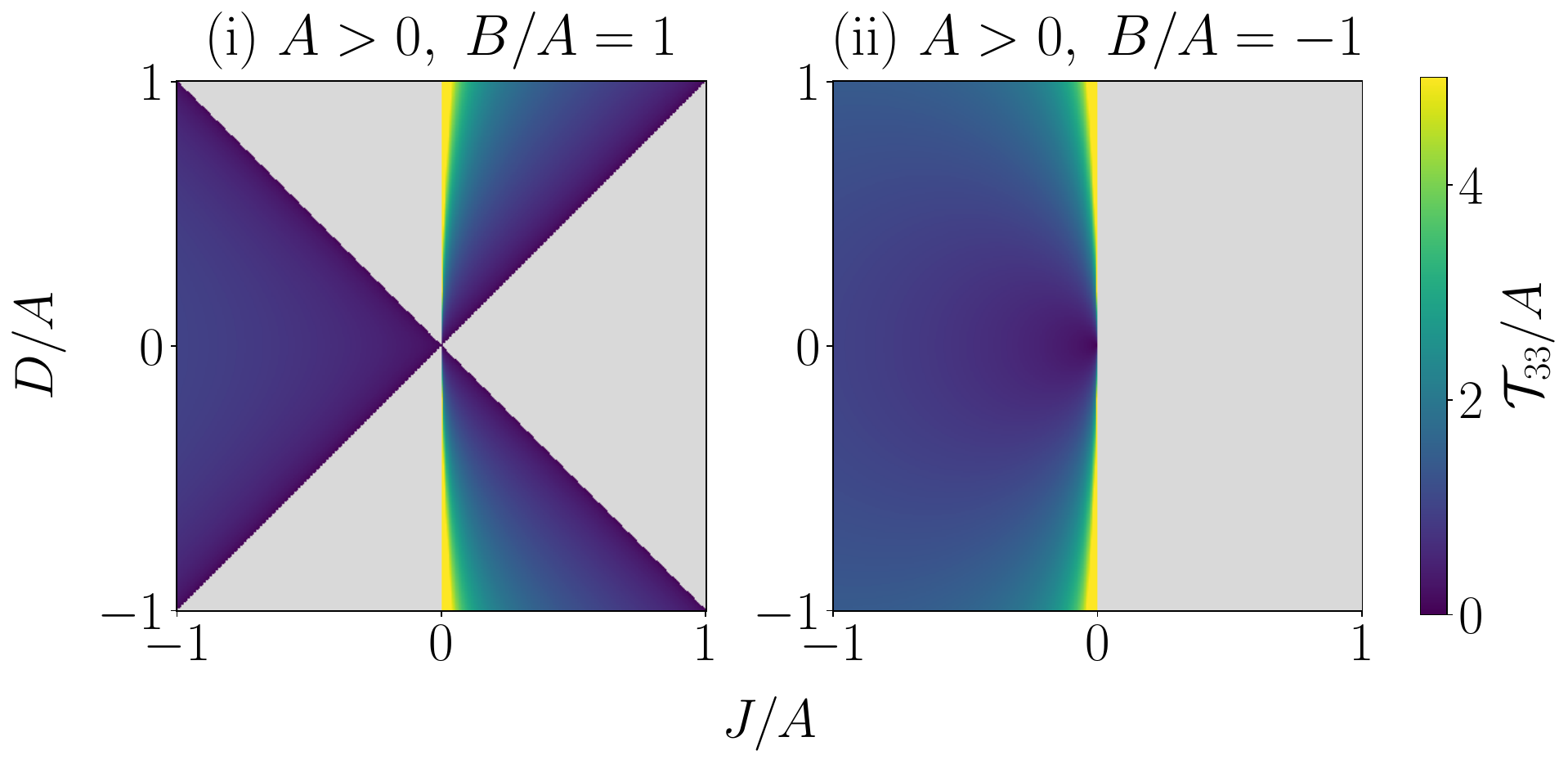}
    \caption{
        Color maps representing the region in which a real solution $\mathcal{T}_{33}$ exists in Eq.~\eqref{T33}.
        The color bar indicates the value of $\mathcal{T}_{33}/A$.
    }
    \label{fig:HeisenbergDMcondition}
\end{figure}

Next, we consider the case where both the Heisenberg and DM couplings are present, given by
\begin{align}
    \mathcal{H} = J \sum_{\mu=x,y,z} S^{\rm L}_{\mu} S^{\rm R}_{\mu} + D (S^{\rm L}_{x} S^{\rm R}_{y} - S^{\rm L}_{y} S^{\rm R}_{x}). \label{H_HDM}
\end{align}
We focus on the configuration shown in Fig.~\ref{fig:setupDM}, where only the leads $ij=$ 11, 22, 33, 12, and 21 are connected.
In this case, Eq.~\eqref{eq:J} implies that only $\mathcal{J}_{\mu\mu}$, $\mathcal{J}_{xy}$, and $\mathcal{J}_{yx}$ have finite values, while all other coupling constants are zero.
The leads 11, 22, and 33 are required to generate the Heisenberg coupling, while leads 21 and 12 generate $\mathcal{J}_{xy}$ and $\mathcal{J}_{yx}$, respectively. Therefore, this setup constitutes the minimal configuration for realizing the Heisenberg~+~DM coupling.
By imposing the target conditions $\mathcal{J}_{\mu\mu} = J$ and $\mathcal{J}_{xy} = -\mathcal{J}_{yx} = D$, we derive the required parameters for the Heisenberg + DM coupling (see Appendix~\ref{App_HDM}):
\begin{align}
    \mathcal{T}_{11} & = -\frac{J}{\widetilde{\chi}_{11,33}\mathcal{T}_{33}},\quad
    \mathcal{T}_{22} = -\frac{J}{\widetilde{\chi}_{22,33}\mathcal{T}_{33}},        \\
    \mathcal{T}_{12} & = -\frac{D}{\widetilde{\chi}_{12,33}\mathcal{T}_{33}},\quad
    \mathcal{T}_{21} = \frac{D}{\widetilde{\chi}_{21,33}\mathcal{T}_{33}},         \\
    \mathcal{T}_{33} & =\sqrt{\frac{BD^{2}-A J^{2}}{J}}, \label{T33}
\end{align}
where $A$ and $B$ are defined as
\begin{align}
    A \coloneqq \frac{\widetilde{\chi}_{11,22}}{\abs{\widetilde{\chi}_{11,33}\widetilde{\chi}_{22,33}}}, \
    B \coloneqq \frac{\widetilde{\chi}_{12,21}}{\abs{\widetilde{\chi}_{12,33}\widetilde{\chi}_{21,33}}}.
\end{align}
Similar to the parameters discussed in Sec.~\ref{sec_XY}, these are determined solely by the properties of the leads and are treated as constants in this section.
Since $\mathcal{T}_{33}$ is a positive real number, $(BD^2-AJ^2)/J > 0$.
The specific conditions for the existence of real solutions, categorized by the signs of $A$ and $B$, are summarized in Table~\ref{tab:HeisenbergDMcondition}.
Furthermore, we visualize the regions satisfying these conditions in Fig.~\ref{fig:HeisenbergDMcondition} by introducing the dimensionless parameters $J/A$ and $D/A$.
In these plots, we assume $|B/A| = 1$ as a representative example.
Note that cases (iii) and (iv) in Table~\ref{tab:HeisenbergDMcondition} are equivalent to cases (ii) and (i), respectively, under the sign reversal of the axes; therefore, we only plot the results for cases (i) and (ii) in Fig.~\ref{fig:HeisenbergDMcondition}.
The colored regions in the figure represent the parameter space where the target Hamiltonian in Eq.~\eqref{H_HDM} is realizable.
The color gradient indicates the required value of the controllable parameter $\mathcal{T}_{33}/A$ to achieve the specific desired values of $(J, D)$.
As seen from Eq.~\eqref{T33}, a small value of $J$ is achieved by choosing a large value of $\mathcal{T}_{33}$.
Furthermore, the slope of the boundary in case (i) is determined by the constant ratio $A/B$.
In contrast, the gray regions signify the regimes where the target Hamiltonian cannot be realized, regardless of how $\mathcal{T}_{33}$ is adjusted, as no real solution exists.
\begin{table}
    \centering
    \caption{The conditions for the existence of a real solution $\mathcal{T}_{33}$ in Eq.~\eqref{T33}.}
    \begin{tabular}{cccc}
        \hline\hline
              & conditions & \( J > 0 \)                     & \( J < 0 \)                           \\
        \hline
        (i)   & $A>0,B>0$  & $\abs{D} > \sqrt{\frac{A}{B}}J$ & $\abs{D} < \sqrt{\frac{A}{B}}\abs{J}$ \\
        (ii)  & $A>0,B<0$  & not exist                       & arbitrary $D$                         \\
        (iii) & $A<0,B>0$  & arbitrary $D$                   & not exist                             \\
        (iv)  & $A<0,B<0$  & $\abs{D} < \sqrt{\frac{A}{B}}J$ & $\abs{D} > \sqrt{\frac{A}{B}}\abs{J}$ \\
        \hline\hline
    \end{tabular}
    \label{tab:HeisenbergDMcondition}
\end{table}

The physical origin of the DM component can be understood as follows.
In the configuration shown in Fig.~\ref{fig:setupDM}, the contribution involving leads $21$ and $33$ generates $\mathcal J_{xy}$, whereas that involving leads $12$ and $33$ generates $\mathcal J_{yx}$. 
Their symmetric and antisymmetric combinations are $\Gamma_{xy}=(\mathcal J_{xy}+\mathcal J_{yx})/2$ and $D=(\mathcal J_{xy}-\mathcal J_{yx})/2$, respectively. 
At the level of the effective spin model, bond-centered inversion symmetry requires $\mathcal J_{xy}=\mathcal J_{yx}$ and therefore forbids $D$.
Microscopically, the inversion operation considered in Appendix~\ref{app:DM_origin} interchanges the corresponding Majorana operators as $\gamma_i^{\rm L}\leftrightarrow\gamma_i^{\rm R}$, exchanges leads $12$ and $21$, and leaves lead $33$ invariant.
Making the inversion-related lead lengths, chemical potentials, tunneling amplitudes, or phase differences inequivalent breaks this inversion symmetry and permits a finite DM component.

Because the target Heisenberg~+~DM Hamiltonian contains no symmetric off-diagonal coupling, the additional condition $\Gamma_{xy}=0$ must be satisfied. 
For the present configuration, this condition is
\begin{align}
\mathcal T_{21}\widetilde{\chi}_{21,33} + \mathcal T_{12}\widetilde{\chi}_{12,33} = 0.
\end{align}
Because $\mathcal T_{12}$ and $\mathcal T_{21}$ are nonnegative, $\widetilde{\chi}_{12,33}$ and $\widetilde{\chi}_{21,33}$ generally must have opposite signs. The required relative sign can be controlled through the lead lengths, chemical potentials, or superconducting phase differences, while the magnitudes of the two contributions can be matched using the gate-controlled tunneling amplitudes. 
Further details of the inversion symmetry and the pure-DM condition are provided in Appendix~\ref{app:DM_origin}.
\section{Spin susceptibility}
\label{4_ss}
In Secs.~\ref{sec_XY} and \ref{sec_HDM}, we focused on modulating the tunneling amplitudes while treating the susceptibilities of the leads as constants.
In this section, we calculate the static spin susceptibility of the leads, defined in Eq.~\eqref{chi_general}, using the tight-binding model given by Eq.~\eqref{H_lead} for two cases: infinite-length leads and finite-length leads.
Based on these results, we finally estimate the magnitude of the RKKY interaction in Sec.~\ref{estimate}.

\subsection{Infinite-length leads}
\label{infinite}
First, we consider infinite-length leads as shown in Fig.~\ref{fig:system_inf}, where the crystal momentum $k$ is a good quantum number.
The energy eigenvalues and eigenstates in Eq.~\eqref{H_lead} for each lead are given by
\begin{gather}
    E_{k}=\varepsilon_0 -2t\cos(k a),
    \\
    \ket|\psi_{k,ij}> = \sum_{r=1}^{N_0} \phi_{k}(r) c_{r,ij}^{\dag} \ket|0>, \\
    \phi_{k}(r) = \frac{1}{\sqrt{N_0}} \exp\qty(ik ra),
\end{gather}
where $k$ is restricted to the first Brillouin zone $-\pi/a \leq k \leq \pi/a$ and $a$ is the lattice constant.
$N_0$ is the number of sites in the lead, which is taken to be sufficiently large ($N_0 \to \infty$).
Hereafter, we set the on-site energy $\varepsilon_0$ to zero.
The coupling sites are located at $x_{\rm L}=1$ and $x_{\rm R}=N$ for all leads.
To obtain an order-of-magnitude estimate of the coupling constant, we assume, for simplicity, that all leads are identical, meaning that they have the same chemical potential and the same length between the junctions.
Then the spin susceptibility in Eq.~\eqref{chi_general} reduces to the isotropic form \cite{RKKY_FiniteSystem_PRB.98.205401}, i.e., not dependent on $ijkl$.
\begin{align}
    \chi
     & =-\qty(\frac{a}{2\pi})^2
    \int_{-\frac{\pi}{a}}^{\frac{\pi}{a}} \dd k \int_{-\frac{\pi}{a}}^{\frac{\pi}{a}} \dd k'\
    \frac{n_{\rm F}(E_{k}) - n_{\rm F}(E_{k'}) }{E_{k}-E_{k'}} \nonumber \\
     & \quad\times \cos\qty[(k - k')(N-1)a],
    \label{infinite_chi}
\end{align}
where $n_{\rm F}(E) = \theta(\varepsilon_{\rm F} - E)$ is the Fermi--Dirac distribution function at zero temperature, and $\varepsilon_{\rm F}$ is the Fermi energy.
In general, when the leads are inequivalent, the spin susceptibility becomes anisotropic and lead dependent, accordingly changing the magnitude and sign of each coupling component. We emphasize, however, that the general formulation of the RKKY interaction in Sec.~\ref{3_RKKY} does not assume equivalent leads and therefore remains applicable in the general case. Consequently, even when the spin susceptibility is lead dependent, the general expressions for the coupling constants and the conditions for realizing the target couplings remain unchanged. The XY and Heisenberg+DM couplings can therefore be realized by adjusting the gate-controlled tunneling amplitudes according to the actual spin susceptibilities.

\begin{figure}
    \centering
    \includegraphics[width=1\linewidth]{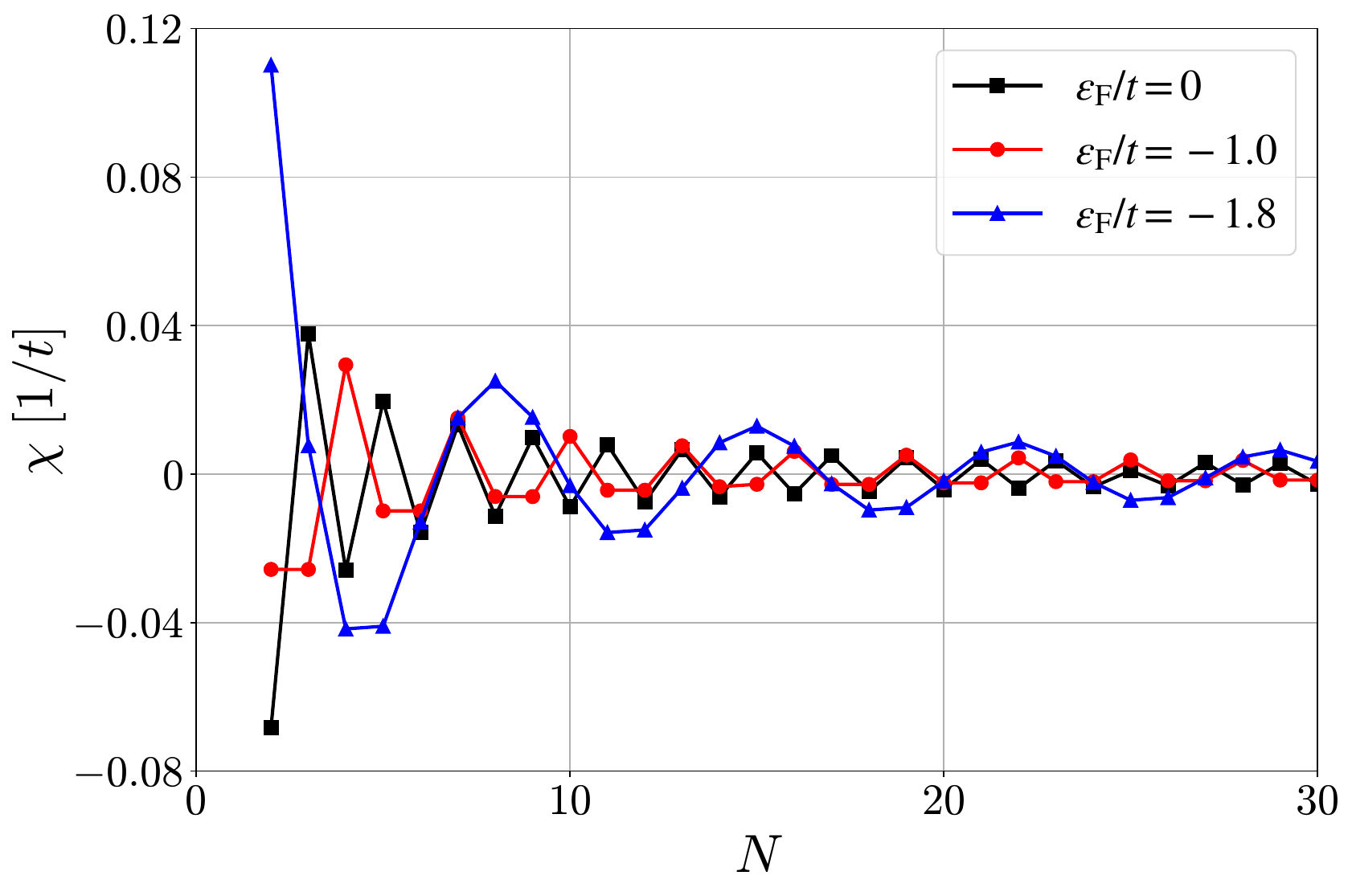}
    \caption{Site $N$ dependence of the isotropic spin susceptibility $\chi$ for infinite length.
        The black, red, and blue lines correspond to the numerical calculation results of the susceptibility for Fermi energies of 0 (at the center of the energy band), $-t$, and $-1.8t$ (around the bottom of the energy band), respectively.}
    \label{fig:infinite-length}
\end{figure}

Figure \ref{fig:infinite-length} displays the calculated isotropic spin susceptibility $\chi$ as a function of the distance $N$ between the two spins $\bm{S}^{\rm L}$ and $\bm{S}^{\rm R}$.
The (black) square, (red) circle, and (blue) triangle symbols denote those for the Fermi energies $\varepsilon_{\rm F}/t=0$, located at the center of the energy band, $-1$, and $-1.8$, located near the bottom of the energy band, respectively.
In all cases, the susceptibility exhibits oscillatory decay.
As $\varepsilon_{\mathrm{F}}/t$ increases from the band bottom, the oscillation period, which corresponds to the Fermi wavelength, becomes shorter.
Therefore, from the perspective of controllability, it is easier to control when the Fermi energy is low.

\subsection{Finite-length leads} \label{finite}
Next, we consider finite-length leads as shown in Fig.~\ref{fig:system_fin} by imposing OBC on the leads Hamiltonian in Eq.~\eqref{H_lead}:
\begin{align}
    \mathcal{H}^{\rm{OBC}}_{\mathrm{lead}}
     & = \sum_{i,j=1}^4
    \Bigg\lbrack \sum_{r=1}^{N} \varepsilon_0 c_{r,ij}^{\dag} c_{r,ij}
    \nonumber           \\
     & \quad
    -t\sum_{r=1}^{N-1} \qty(c_{r,ij}^{\dag} c_{r+1,ij} +\mathrm{h.c.} )
    \Bigg\rbrack,
\end{align}
where we assume $x_{\rm L} = 1$ and $x_{\rm R} = N$ similar to the infinite-length case.
Then, the energy eigenvalues and eigenstates are given by
\begin{align}
    E_{\lambda}
                      & =\varepsilon_0 -2t\cos\qty(\frac{\lambda\pi}{N+1})
    \quad (\lambda=1,\cdots,N),
    \\
    \ket|\psi_{\lambda}>
                      & = \sum_{r=1}^{N}
    \phi_{\lambda}(r) c_{r,ij}^{\dag} \ket|0>,
    \\
    \phi_{\lambda}(r) & =
    \frac{1}{\sqrt{\sum_{k=1}^{N} \sin^2
            \qty(\frac{\lambda\pi k}{N+1})}}
    \sin \qty(\frac{\lambda\pi r}{N+1}),
\end{align}
where $\lambda$ is the label of the eigenstate.
This eigenfunction indeed satisfies the OBC $\phi_{\lambda}(0)=\phi_{\lambda}(N+1)=0$.

\begin{figure}
    \centering
    \includegraphics[width=1\linewidth]{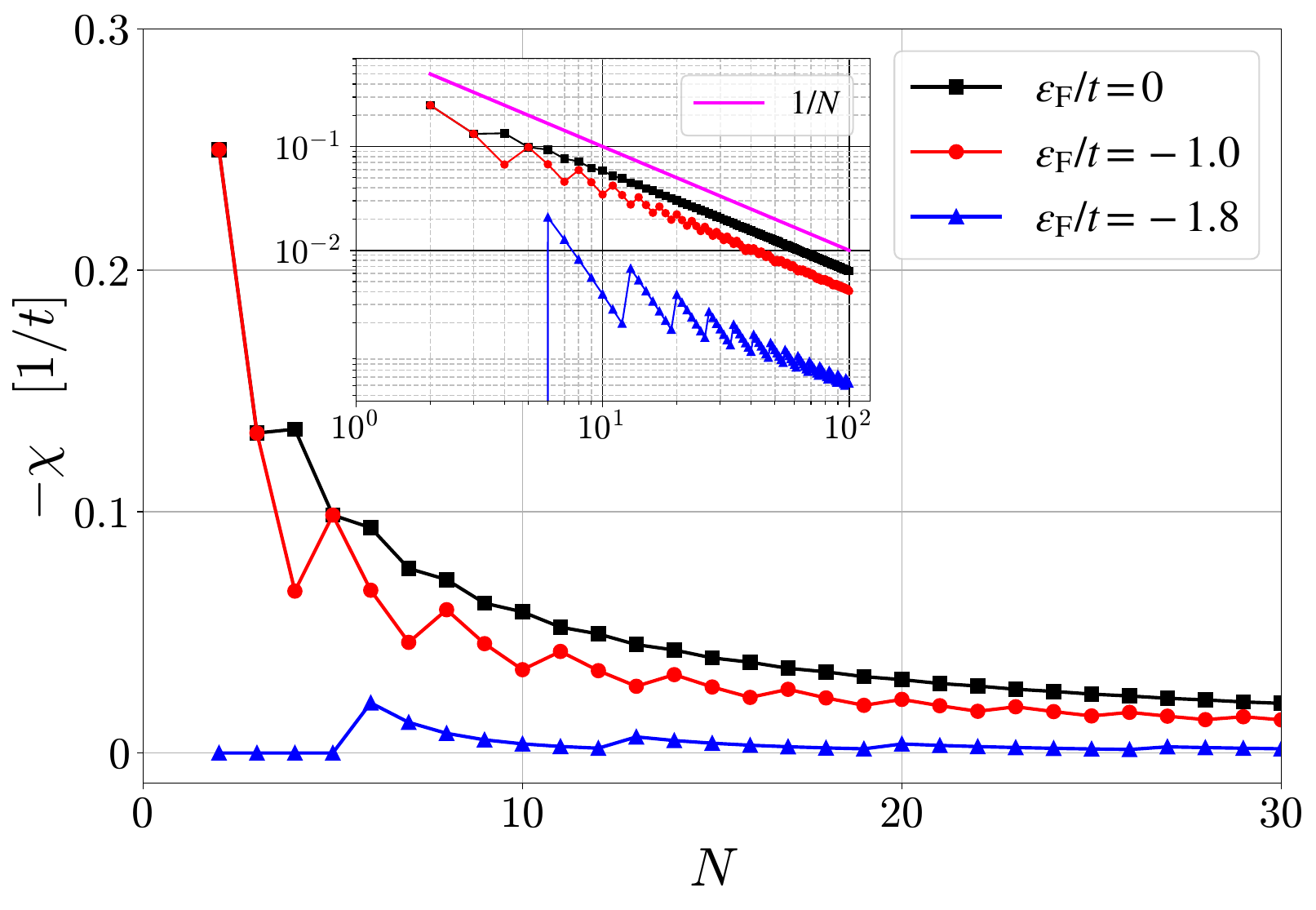}
    \caption{
        The isotropic spin susceptibility for finite length cases at three different Fermi energies: 0, $-t$, and $-1.8t$.
        The inset is a double-logarithmic plot that shows the decay is approximately as $1/N$.}
    \label{fig:finite-length}
\end{figure}

The isotropic spin susceptibility $\chi$ in the finite-length case is calculated analogously to Eq.~\eqref{infinite_chi},
by replacing the integration over the wavenumber $k$ with a summation over the discrete quantum number $\lambda$:
\begin{align}
    \chi
     & = -
    \sum_{\lambda,\lambda'=1 }^{N}
    \frac{n_{\rm F}(E_\lambda) - n_{\rm F}(E_{\lambda'})}{E_\lambda-E_{\lambda'}}
    \nonumber      \\
     & \quad\times
    \phi_{\lambda}(N) \phi^{\ast}_{\lambda}(1)\phi_{\lambda'}(1) \phi^{\ast}_{\lambda'}(N).
\end{align}
Figure~\ref{fig:finite-length} shows the $N$-dependence of the spin susceptibility for the finite-length case.
It decays approximately as $1/N$ (see the inset).
A significant difference from the infinite-length case is the absence of sign-changing oscillations.
Consequently, from the perspective of controllability, this monotonic behavior is advantageous when one wishes to avoid unintended sign reversals.

\subsection{Order estimate of the RKKY interaction}
\label{estimate}
In this section, we estimate the order of magnitude of the RKKY coupling constant in Eq.~\eqref{HRKKY2}.
We assume that the hopping energy $t$ within the leads is of the order of 1 eV, and the Fermi energy is $\varepsilon_{\mathrm{F}} = -1.8 t$.
For the finite-length leads, the magnitude of the isotropic spin susceptibility at a length $l \sim 10\, \mathrm{nm}$ ($N=30$, $a=3.3$ \AA) between two MCBs is estimated to be at most $\abs{\chi} \sim 0.1\, /\mathrm{eV}$.
The charging energy $E_{\rm c}$ of an MCB is on the order of $1\, \mathrm{meV}$ \cite{MCB2016}.
While the perturbative expansion requires $t_{\rm tun} \ll E_{\rm c}$, we nevertheless adopt a rather large value of $t_{\rm tun} = E_{\rm c}/2$ within the acceptable range to evaluate the possible scale of the RKKY coupling.
Here, $t_{\rm tun}$ denotes the typical magnitude of a tunneling amplitude $\tilde t_{ij}$ between a Majorana and an electron in the lead.
Accordingly, the magnitude of the uniform coupling $\mathcal{T}_{ij}$ in Eq.~\eqref{Jij} is estimated as $\mathcal{T} \approx 4t_{\rm tun}^2/E_{\rm c}= 1\, \mathrm{meV}$.
Finally, the strength of the RKKY interaction is estimated as $\abs{\mathcal{J}_{\rm RKKY}} \approx \mathcal{T}^2 \abs{\chi} \sim 0.1 \, \mu\mathrm{eV}$.
This energy scale corresponds to a temperature of approximately $1\, \mathrm{mK}$, which is within the reach of current experimental techniques.

We clarify the regime of validity of the perturbative RKKY description. The RKKY energy scale estimated above should be compared with the Kondo temperature of an isolated MCB.
The Kondo scale is exponentially small in the weak-coupling regime, $T_{\rm K}\sim D_0 e^{-{1}/{\rho \mathcal T}}$, where $D_0$ is the high-energy cutoff of the effective Kondo model and $\rho$ is the density of states of the leads at the Fermi level.
Thus, the present RKKY-based spin Hamiltonian applies when
$
    |\mathcal J_{\rm RKKY}|
    \gg
    T_{\rm K}$.
The crossover to the Kondo-dominated regime is expected around
$
    |\mathcal J_{\rm RKKY}|
    \sim
    T_{\rm K}.
$
A quantitative treatment of this crossover requires a nonperturbative analysis and is left for future work.
\section{Summary}
\label{5_Summary}
We have theoretically demonstrated that connecting MCBs via multiple normal metal leads provides a versatile framework for engineering spin couplings described by quadratic spin Hamiltonians.
Our central finding is that the wiring patterns of the leads serve as a crucial degree of freedom for determining the form of the resulting RKKY interaction.
Our method realizes not only the standard Heisenberg coupling but also the XY coupling and asymmetric terms such as the DM interaction, which is tuned by modulating the tunneling amplitude via gate voltages, adjusting the length or chemical potential of the leads, and inducing a phase difference between the superconducting order parameters by driving a Josephson current.

The design established here for a two-MCB system can be extended to larger networks.
By arranging MCBs in one- or two-dimensional lattices and engineering the wiring patterns accordingly, our approach provides a versatile platform for exploring a diverse range of quantum spin systems.
When we connect only the nearest-neighbor MCBs, the lowest-order RKKY interaction generates solely the nearest-neighbor coupling $J_1$. On the other hand, next-nearest-neighbor couplings $J_2$ can also be induced either by considering higher-order perturbation effects or by extending the lead connections to the next-nearest-neighbor sites.
However, the main contribution to $J_2$ arising from higher-order corrections is of the fourth order in $\mathcal{T}$, and its relative magnitude is estimated to be $|J_2|/|J_1|\sim(\mathcal{T}|\chi|)^2\ll1$. Therefore, $J_2$ is sufficiently small compared to $J_1$ and can be neglected within the range of the lowest-order RKKY interaction.

Finally, we comment on the validity of the perturbative approach used in this work.
Our derivation of the RKKY interaction assumes a weak coupling regime between the Majoranas and the conduction electrons.
However, in the strong coupling regime, it is known that MCB systems can exhibit the topological Kondo effect, characterized by non-Fermi liquid behavior \cite{beriTopologicalKondoEffect2012,MajoranaKleinHybridizationTopological2013,Multiterminal_Altland,Altland_2014,Multichannel_Altland,buccheriThermodynamicsTopologicalKondo2015,herviouManyterminalMajoranaIsland2016,liMultichannelTopologicalKondo2023}.
Consequently, a competition between the RKKY interaction and the topological Kondo effect is expected, analogous to the Doniach phase diagram in heavy fermion systems \cite{Doniach1977}.
Exploring the phase diagram and the crossover between these regimes remains an intriguing and important task for future research.
\begin{acknowledgments}
    M.T. was supported by JST SPRING (Grant No.~JPMJSP2125).
    S.H. was supported by JSPS KAKENHI (Grant Nos.~23K17668, 24K00578, and 24K00583) and the Grant-in-Aid for Transformative Research Areas (A) “Correlation Design Science” (Grant No.~JP25H01249).
    A.Y. was supported by JSPS KAKENHI (Grant Nos.~JP24H00853 and JP25K07224).
\end{acknowledgments}
\section*{Data Availability}
The data that support the findings of this article are publicly available~\cite{data}.

\appendix

\section{RKKY coupling}\label{App_RKKY}
Here, we derive the RKKY coupling Eq.~\eqref{Jij}.
First, we define the Matsubara Green's function as
\begin{align}
  G_{ab,cd}(x_{\rm L},x_{\rm R};\tau_1-\tau_2) 
  &= -\expval{T_\tau c_{x_{\rm L},ab}(\tau_1) c_{x_{\rm R},cd}^{\dagger}(\tau_2)} \nonumber\\
  &=G_{ab}(x_{\rm L},x_{\rm R};\tau_1-\tau_2)\delta_{ac}\delta_{bd}.
\end{align}
This Green's function involves two leads, $ab$ and $cd$, but because there is no interaction between different leads, it simplifies as shown in the second line.
Using this Green's function, Eq.~\eqref{Jijij} can be rewritten as follows:
\begin{widetext}
\begin{align}
    \mathcal J_{ij,i'j'}
    &=T^{\rm L}_{ij',ji'} T^{\rm R}_{ji',ij'} \qty[e^{i(\Delta\phi_{ij'}-\Delta\phi_{ji'})} \chi_{ij',ji'}+e^{-i(\Delta\phi_{ij'}-\Delta\phi_{ji'})} \chi_{ji',ij'}] \nonumber\\
    &\quad-T^{\rm L}_{ii',jj'} T^{\rm R}_{ii',jj'} \qty[e^{i(\Delta\phi_{ii'}-\Delta\phi_{jj'})} \chi_{ii',jj'}+e^{-i(\Delta\phi_{ii'}-\Delta\phi_{jj'})} \chi_{jj',ii'}], \label{App_Jij}
\end{align}
where
\begin{align}
    \chi_{ab,cd}= -T\int_0^{1/T} \dd \tau_1 \int_0^{1/T} \dd \tau_2\
    G_{ab}(x_{\rm L},x_{\rm R};\tau_1-\tau_2)G_{cd}(x_{\rm R},x_{\rm L};\tau_2-\tau_1) \label{chi_abcd}.
\end{align}
\end{widetext}
In addition, we perform a Fourier transform of the Matsubara Green's function into Matsubara frequency space
\begin{align}
    G_{ab}(x_{\rm L},x_{\rm R};\tau)=T\sum_l G_{ab}(x_{\rm L},x_{\rm R};i\omega_l)e^{-i\omega_l\tau},
\end{align}
where $\omega_l=(2l+1)\pi T$ and $l$ is an integer.
Furthermore, by rewriting the Matsubara frequency summation as a contour integral in the complex plane, Eq.~\eqref{chi_abcd} reduces to
\begin{align}
  \chi_{ab,cd}
  &= \int_{-\infty}^\infty\frac{\dd \varepsilon}{2\pi i} n_{\rm F}(\varepsilon) 
  \Big\lbrack G_{ab}^{\rm R}(x_{\rm L},x_{\rm R};\varepsilon)G_{cd}^{\rm R}(x_{\rm R},x_{\rm L};\varepsilon)\nonumber\\
  &\quad-G_{ab}^{\rm A}(x_{\rm L},x_{\rm R};\varepsilon)G_{cd}^{\rm A}(x_{\rm R},x_{\rm L};\varepsilon) \Big\rbrack, \label{chi_abcd2}
\end{align} 
where $n_{\rm F}(\varepsilon) = (e^{\varepsilon/T}+1)^{-1}$ is the Fermi-Dirac distribution function and the superscripts R/A denote the retarded/advanced Green's function.
Using the relationship $[G_{ab}^{\rm R}(x_{\rm L},x_{\rm R};\varepsilon)]^\ast=G_{ab}^{\rm A}(x_{\rm R},x_{\rm L};\varepsilon)$, we obtain the relationship $\chi^\ast_{ab,cd}=\chi_{cd,ab}$ from Eq.~\eqref{chi_abcd2}.
Applying this relationship to rewrite Eq.~\eqref{App_Jij}, we obtain Eq.~\eqref{Jij}.

In particular, when the system has inversion symmetry, or when the lead index of the Green’s function is irrelevant, Eq.~\eqref{chi_abcd2}  simplifies and can be written as follows at zero temperature:
\begin{align}
  \chi_{ab,cd}
  &=\frac{1}{\pi}\int_{-\infty}^{\varepsilon_{\rm F}}\dd \varepsilon\
  \mathrm{Im}\left[G_{ab}^{\rm R}(x_{\rm L},x_{\rm R};\varepsilon)G_{cd}^{\rm R}(x_{\rm R},x_{\rm L};\varepsilon)\right].
\end{align}

\renewcommand{\thefigure}{B\arabic{figure}}
\renewcommand{\theHfigure}{B\arabic{figure}}
\setcounter{figure}{0}
\section{XY coupling} \label{App_XY}
In Sec.~\ref{sec_XY}, we consider only a particularly simple case. 
Here, we present the general conditions under which the XY coupling is realized.
First, by eliminating $\tilde{\mathcal{T}}_{22}$ from Eqs.~\eqref{XY_Jx} and \eqref{XY_Jy}, we obtain 
\begin{align}
    \tilde{J}=\qty[\qty(\frac{\tilde{J}}{\tilde{A}_y \tilde{\mathcal{T}}_{11}^2-\tilde{B}_y})^2 - \tilde{B}_x]\tilde{\mathcal{T}}_{11}. \label{XYcond}
\end{align}
Here, the following two conditions are imposed: $\tilde{A}_y \tilde{\mathcal{T}}_{11}^2-\tilde{B}_y \neq 0$, and $\mathrm{sgn}(\tilde{J})=\mathrm{sgn}(\tilde{A}_y \tilde{\mathcal{T}}_{11}^2-\tilde{B}_y)$. 
The latter is equivalent to $\tilde{\mathcal{T}}_ {22}>0$.
Solving $\tilde{J}$ as a function of $\tilde{\mathcal{T}}_{11}$, we obtain two solutions
\begin{align}
    \tilde J_{\pm}&=
    \frac{(\tilde{A}_y \tilde{\mathcal{T}}_{11}^2 - \tilde{B}_y)^2}{2\tilde{\mathcal{T}}_{11}}
    \qty(1\pm \sqrt{ 1 + \frac{4 \tilde B_x \tilde{\mathcal{T}}_{11}^2}{(\tilde A_y \tilde{\mathcal T}_{11}^2 - \tilde B_y)^2} }),
    \label{XY_JT11}
\end{align}
where the subscript $\pm$ represents the two branches for $\tilde J$.
Note that for $\tilde J_{\pm}$ to have real solutions, the expression inside the square root must be nonnegative.
Then, $\tilde{\mathcal{T}}_{22}$ is given for these two branches as 
\begin{align}
    \tilde{\mathcal{T}}_{22,\pm}&=
    \frac{\tilde J_{\pm}}{\tilde A_y \tilde{\mathcal T}_{11}^2 - \tilde B_y},
    \label{XY_JT22}
\end{align}
respectively.

\begin{figure}
    \centering
    \includegraphics[width=1\linewidth]{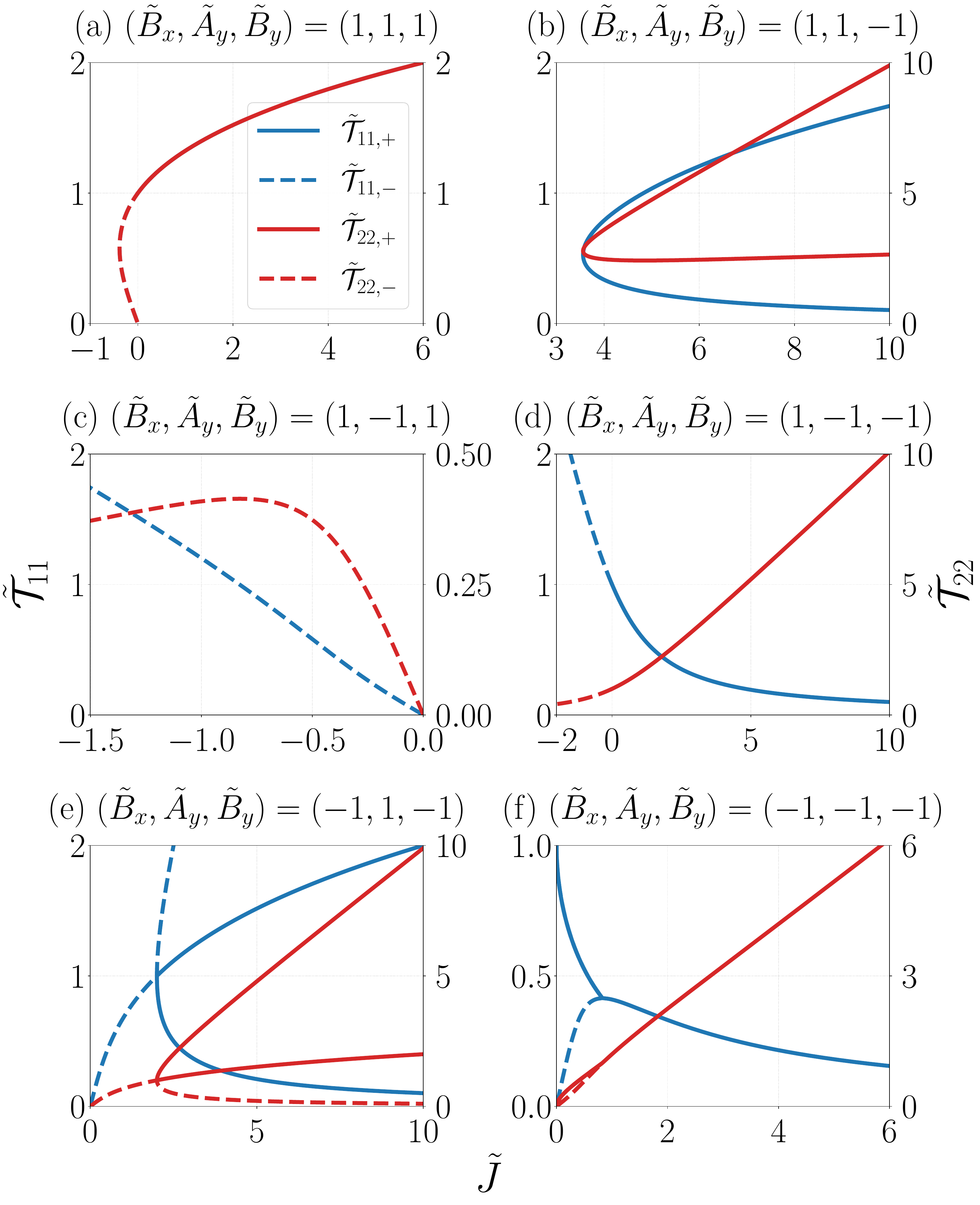}
    \caption{Plots showing $\tilde{\mathcal{T}}_{11}$ (left vertical axis, blue line) and $\tilde{\mathcal{T}}_{22}$ (right vertical axis, red line) as functions of $\tilde J$ for the case $|\tilde B_x| = |\tilde A_y| = |\tilde B_y| = 1$.
    The solid and dashed lines correspond to the $\tilde J_{+}$ and $\tilde J_{-}$ branches, respectively.}
    \label{fig:XY}
\end{figure}

Figure~\ref{fig:XY} shows $\tilde{\mathcal T}_{11}$ and $\tilde{\mathcal T}_{22}$ as functions of $\tilde J$, obtained from Eqs.~\eqref{XY_JT11} and \eqref{XY_JT22}, respectively.
Here, since the constants $A_i$ and $B_i$ are of the same order of magnitude, we set $|\tilde B_x| = |\tilde A_y| = |\tilde B_y| = 1$.
The blue line represents $\mathcal{T}_{11}$, the red line represents $\mathcal{T}_{22}$, and the solid and dashed lines correspond to the $\tilde J_\pm$ branches, respectively.
The cases of $(\tilde B_x,\tilde A_y,\tilde B_y)=(-1, 1, 1)$ and $(-1, -1, 1)$ are not plotted because no real solutions exist, i.e., the XY coupling cannot be realized.
Note that in case (a), the blue line and the red line are completely overlapping.
In cases (e) and (f), there can be two possible values of $\tilde J$ for a single value of $\tilde{\mathcal T}_{11}$. 
However, since $\tilde{\mathcal T}_{22}$ also has two branches corresponding to the two branches of $\tilde J_{\pm}$, specifying the pair $(\tilde{\mathcal T}_{11}, \tilde{\mathcal T}_{22})$ uniquely determines $\tilde J$.
Among the six panels, the most convenient case is (d), where the value of $\tilde J$ can be controlled relatively freely, including both positive and negative values, by modulating $\tilde{\mathcal T}_{11}$ and $\mathcal{T}_{22}$ with the gate voltage.

To summarize, first, from Eq.~\eqref{XY_JT11}, the value of variable $\mathcal{T}_{11}$ is determined for the desired $J$.
Next, from Eq.~\eqref{XY_JT22}, the value of variable $\mathcal{T}_{22}$ is also uniquely determined.
Finally, by determining the remaining value of $\mathcal{T}_{33}$ from Eq.~\eqref{XY_T33}, we can realize the XY coupling with the desired coupling constant $J$.

\section{Heisenberg + DM coupling} \label{App_HDM}
Here, we derive the conditions for the realization of the Heisenberg + DM coupling.
Now we consider the configuration shown in Fig.~\ref{fig:setupDM}, where only the leads 11, 22, 33, 12, and 21 are connected.
In this case, according to Eq.~\eqref{eq:J}, the coupling constants satisfy 
\begin{align}
\mathcal{J}_{xx}
&= - \mathcal{T}_{22}\mathcal{T}_{33}\widetilde{\chi}_{22,33}, \\
\mathcal{J}_{yy} 
&= - \mathcal{T}_{11}\mathcal{T}_{33} \widetilde{\chi}_{11,33}, \\
\mathcal{J}_{zz} 
&= \mathcal{T}_{12}\mathcal{T}_{21} \widetilde{\chi}_{12,21} - \mathcal{T}_{11}\mathcal{T}_{22} \widetilde{\chi}_{11,22},  \\
\mathcal{J}_{xy} 
&= \mathcal{T}_{21}\mathcal{T}_{33} \widetilde{\chi}_{21,33}, \label{eq:App_Jxy} \\
\mathcal{J}_{yx} 
&= \mathcal{T}_{12}\mathcal{T}_{33} \widetilde{\chi}_{12,33} \label{eq:App_Jyx} ,
\end{align}
and all other $\mathcal{J}_{ij}$ are zero.  

First, from the condition $\mathcal{J}_{xx} = \mathcal{J}_{yy} = J$, we obtain
\begin{align}
    -\widetilde{\chi}_{22,33} \mathcal{T}_{22} 
    =-\widetilde{\chi}_{11,33} \mathcal{T}_{11}
    =\frac{J}{\mathcal{T}_{33}}. \label{xx_yy_J}
\end{align}
Since $\mathcal{T}_{ij}\geq 0$, the following condition holds: $\mathrm{sgn}(J)=-\mathrm{sgn}(\widetilde{\chi}_{11,33})=-\mathrm{sgn}(\widetilde{\chi}_{22,33})$.
It thus follows that $\widetilde{\chi}_{11,33}$ and $\widetilde{\chi}_{22,33}$ must always have the same sign.

Next, from the condition $\mathcal{J}_{xy} = -\mathcal{J}_{yx} = D$, we obtain
\begin{align}
    \widetilde{\chi}_{21,33} \mathcal{T}_{21} 
    =-\widetilde{\chi}_{12,33} \mathcal{T}_{12}
    = \frac{D}{\mathcal{T}_{33}}. \label{xy_yx_D}
\end{align}
Therefore, the following condition is imposed: $\mathrm{sgn}(D)=-\mathrm{sgn}(\widetilde{\chi}_{12,33})=\mathrm{sgn}(\widetilde{\chi}_{21,33})$.
It thus follows that $\widetilde{\chi}_{12,33}$ and $\widetilde{\chi}_{21,33}$ must always have opposite signs.

Finally, by substituting Eqs.~\eqref{xx_yy_J} and \eqref{xy_yx_D} into the condition $\mathcal{J}_{zz} = J$, we obtain the conditions for $J$ and $D$ to realize the Heisenberg + DM coupling as follows:
\begin{align}
    \begin{cases}
    A {J}^2 + \mathcal T_{33}^2 J - B{D}^2 = 0, \label{HDM} \\
    \mathrm{sgn}(J)=-\mathrm{sgn}(\widetilde{\chi}_{11,33})=-\mathrm{sgn}(\widetilde{\chi}_{22,33}),\\
    \mathrm{sgn}(D)=-\mathrm{sgn}(\widetilde{\chi}_{12,33})=\mathrm{sgn}(\widetilde{\chi}_{21,33}).
\end{cases}
\end{align}
It then follows that the externally controllable parameter $\mathcal{T}_{33}$ can be written as a function of the target parameters $J$ and $D$ as Eq.~\eqref{T33}.

\section{Inversion symmetry and the condition for a pure DM contribution}
\label{app:DM_origin}

In this Appendix, we explain how the cross-label leads generate the DM component, clarify its relation to inversion symmetry, and derive the condition for obtaining a pure DM contribution.

The off-diagonal coupling can be decomposed as
\begin{align}
\mathcal H_{\mathrm{off}}
={}&
\Gamma_{xy}
\left(
S_x^{\rm L}S_y^{\rm R}
+
S_y^{\rm L}S_x^{\rm R}
\right)
\nonumber\\
&+
D_z
\left(
S_x^{\rm L}S_y^{\rm R}
-
S_y^{\rm L}S_x^{\rm R}
\right),
\label{eq:offdiagonal_decomposition}
\end{align}
where
\begin{align}
\Gamma_{xy}
&=
\frac{\mathcal J_{xy}+\mathcal J_{yx}}{2},
&
D_z
&=
\frac{\mathcal J_{xy}-\mathcal J_{yx}}{2}.
\label{eq:Gamma_D_definition}
\end{align}
Here, \(\Gamma_{xy}\) is the symmetric off-diagonal coupling coefficient, while \(D_z\) is the coefficient of
\[
S_x^{\rm L}S_y^{\rm R}
-
S_y^{\rm L}S_x^{\rm R}
=
(\bm S^{\rm L}\times\bm S^{\rm R})_z.
\]
Thus, a finite DM component requires
\(\mathcal J_{xy}\neq\mathcal J_{yx}\).

At the level of the effective spin model, bond-centered inversion interchanges the two MCBs:
\begin{align}
\mathcal I S_\mu^{\rm L}\mathcal I^{-1}
&=
S_\mu^{\rm R},
&
\mathcal I S_\mu^{\rm R}\mathcal I^{-1}
&=
S_\mu^{\rm L}.
\label{eq:spin_inversion}
\end{align}
This operation interchanges the two off-diagonal coupling terms. Inversion symmetry therefore requires
\(\mathcal J_{xy}=\mathcal J_{yx}\) and hence forbids \(D_z\).

Microscopically, different Majorana bilinears represent different effective spin components. Consequently, the contribution involving leads \(21\) and \(33\) generates \(\mathcal J_{xy}\), whereas that involving leads \(12\) and \(33\) generates \(\mathcal J_{yx}\), as shown in Eqs.~\eqref{eq:App_Jxy} and \eqref{eq:App_Jyx}. Their symmetric and antisymmetric combinations are
\begin{align}
\Gamma_{xy}
&=
\frac{\mathcal T_{33}}{2}
\left(
\mathcal T_{21}\widetilde{\chi}_{21,33}
+
\mathcal T_{12}\widetilde{\chi}_{12,33}
\right),
\label{eq:Gamma_xy}
\\
D_z
&=
\frac{\mathcal T_{33}}{2}
\left(
\mathcal T_{21}\widetilde{\chi}_{21,33}
-
\mathcal T_{12}\widetilde{\chi}_{12,33}
\right).
\label{eq:Dz_microscopic}
\end{align}

We now consider a microscopic inversion operation that preserves the Majorana labels while interchanging the two MCBs:
\begin{align}
\mathcal I\gamma_i^{\rm L}\mathcal I^{-1}
&=
\gamma_i^{\rm R},
&
\mathcal I\gamma_i^{\rm R}\mathcal I^{-1}
&=
\gamma_i^{\rm L},
&
\mathcal I c_{r,ij}\mathcal I^{-1}
&=
c_{\bar r,ji}.
\label{eq:microscopic_inversion}
\end{align}
Here, \(\bar r\) denotes the spatially inverted coordinate along the lead. Since lead \(12\) connects \(\gamma_1^{\rm L}\) and \(\gamma_2^{\rm R}\), inversion maps it onto lead \(21\), which connects \(\gamma_2^{\rm L}\) and \(\gamma_1^{\rm R}\). Lead \(33\) is mapped onto itself.

For the microscopic Hamiltonian to possess this inversion symmetry, leads \(12\) and \(21\) must therefore be equivalent under inversion. 
In the lead model adopted here, this requires the two leads to have the same length and chemical potential, with their contact positions related by inversion.
Furthermore, the complex tunneling amplitudes must also satisfy
\begin{align}
\widetilde t_{12}^{\rm L}
&=
\widetilde t_{21}^{\rm R},
&
\widetilde t_{12}^{\rm R}
&=
\widetilde t_{21}^{\rm L}.
\label{eq:tunneling_inversion}
\end{align}
These conditions make the two crossed-channel contributions equivalent:
\begin{align}
\mathcal T_{21}\widetilde{\chi}_{21,33}
=
\mathcal T_{12}\widetilde{\chi}_{12,33}.
\label{eq:inversion_exchange_condition}
\end{align}
It follows that \(\mathcal J_{xy}=\mathcal J_{yx}\) and \(D_z=0\).
Conversely, making the inversion-related lead lengths, chemical potentials, tunneling amplitudes, or phase differences inequivalent breaks this symmetry and permits the two contributions to become unequal.

The target Heisenberg~+~DM Hamiltonian contains no symmetric off-diagonal coupling. A pure DM contribution therefore requires \(\Gamma_{xy}=0\), or equivalently,
\begin{align}
\mathcal T_{21}\widetilde{\chi}_{21,33}
+
\mathcal T_{12}\widetilde{\chi}_{12,33}
=0.
\label{eq:pure_DM_condition}
\end{align}
Because \(\mathcal T_{12},\mathcal T_{21}\geq0\), this condition generally requires
\(\widetilde{\chi}_{12,33}\) and
\(\widetilde{\chi}_{21,33}\) to have opposite signs.
Their relative signs can be controlled through the lead lengths, chemical potentials, or superconducting phase differences, while the magnitudes of the two contributions can be matched by tuning the gate-controlled tunneling amplitudes.

\bibliography{ref}
\end{document}